\documentclass{svjour3} 
\usepackage{graphicx}        
\smartqed  
\begin{document}
\title{On the telescopes in the paintings of J. Brueghel the Elder}
\author{Pierluigi Selvelli$^1$        \and
        Paolo Molaro$^1$}
\institute{\at$^1$ INAF-Osservatorio Astronomico di Trieste,  Via Tiepolo 11,
I-34143, Trieste, Italy\\
              Tel.: 39-040-3199218\\
              Fax: 39-040-309418\\
              \email{selvelli@oats.inaf.it, molaro@oats.inaf.it}          
}

\date{Received: date / Accepted: date}

\maketitle

\begin{abstract}

 We have  investigated the nature and the origin of the  telescopes
depicted in three paintings of J. Bruegel the Elder completed between 1609 and
1618.
  The   "tube" that appears in the painting dated 1608-1612  represents  a
very early dutch
spyglass, tentatively  attributable to Sacharias Janssen or Lipperhey, prior
to those
made by Galileo, while the two  instruments made of several draw-tubes 
which appear in the  two paintings of 1617 and 1618 are quite sophisticated
and may represent early
examples of Keplerian telescopes.

\end{abstract}

\section{Introduction}
\label{intro}

 The presence of  spyglasses and  other astronomical
instruments  characterizes  three paintings of J. Brueghel the Elder
(1568-1625) composed between 1609 and 1618.  
The paintings were drawn while the artist was appointed as a  painter at
the
court of  the Archduke Albert VII of Habsburg (1559-1621), spanish
Governor of the catholic part of the Netherlands, and of his spouse, the
Archduchess
Isabella Clara Eugenia, "Infanta" of Spain and daughter of Felipe 
II.

 The three paintings  can be 
considered as a  high quality, realistic  photographies and are  rich in  
naturalistic details.  This allows a  detailed examination of the 
instruments therein depicted and an investigation of their   nature.

\begin{figure}
  \includegraphics[width=0.94\textwidth]{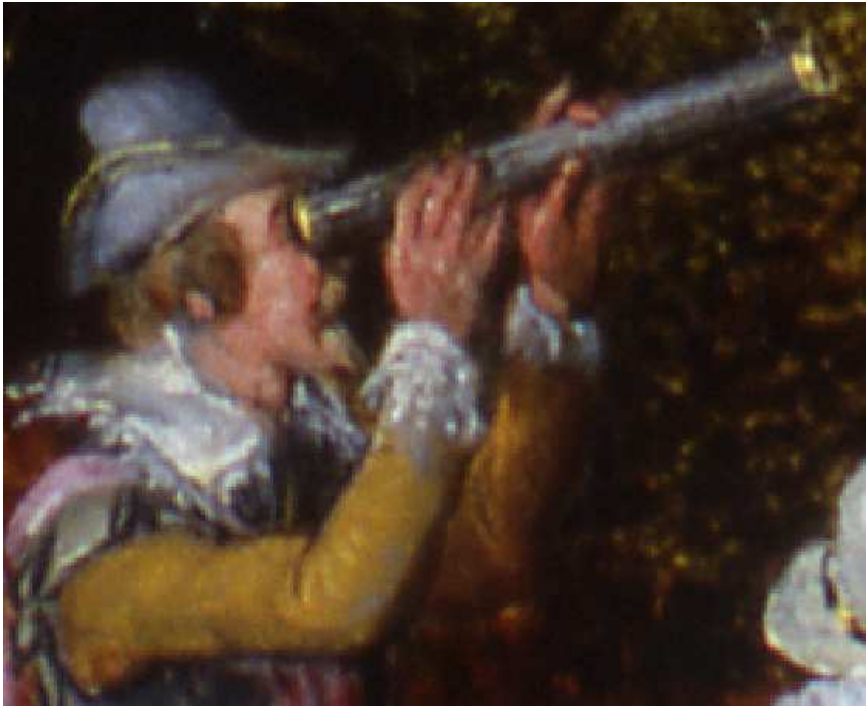}
\caption{A detail of the painting "Extensive Landscape with View of the
Castle of Mariemont" by J. Brueghel the Elder, ca. 1609-1612.  Virginia
Museum of Fine Arts, Richmond. The Adolph D. and Wilkins C. Williams 
Fund. Photo: Ron Jennings. 
The actual length of the spyglass  is estimated to be  about 40 cm. }
\label{fig:1}     
\end{figure}

\section{The painting with the early spyglass} 

The painting "Extensive Landscape with View of the Castle of Mariemont"  
is conserved at the Virginia Museum of Fine Arts (VMFA)  in Richmond, VA,
USA.  The   Mariemont Castle, located near Bruxelles,  was under reformation
for
several years, and a study of the development of the building,  by
comparison with 
paintings of ascertained dates, has indicated that 
 the date of   the painting completion was  in the years  1608 - 1611 .
In the central part of the painting one can see the important detail of
the archduke Albert VII looking through a spyglass.
The instrument has a cylindrical shape and appears as  metallic, probably
made of tin,   with two  gilded rings   on either side.  (Fig. 1). By
comparison
with other figures and objects one  can estimate that its  length  is 
about 40-45
cm and its diameter of about 5 cm.

To our knowledge,  this painting  represents the most ancient reproduction
of   spyglasses and we  guess    that the depicted instrument actually
represents  
one of the  first spyglasses ever built.  
We know from 
Daniello Antonini, who was serving with the Archduke
in Brussels, that Albert VII owned copies of early spyglasses. Antonini,  
on September 2, 1611, wrote a letter to his former master Galileo (Bibl. Naz.
Fir., Mss. 
Gal., P. VI, T. VIII, car.37)
informing him   that the Archduke had obtained some spyglasses from the "first
inventor", 
although of
lower quality in comparison  to those owned by Galileo.
Thus, it is likely that  the "tube" held by the Archduke represents one
of these spyglasses mentioned by Antonini.
Concerning the manufacturer,  it could be either Sacharias  Janssen  or 
Hans Lipperhey, both considered  "fathers" of the telescope.  
According to Pierre Borel (Van Helden, 1977), Sacharias Janssen made some
instruments measuring about 40 cm  in length, and offered the
best ones to Prince Maurice of Nassau and the Archduke Albert.
Alternatively,  according to Antonius Maria Schyrley de Rheita, 
the Marquis Ambrogio Spinola, Commander of the Spanish Army in the
Flanders, bought a spyglass  in The
Hague near the end of 1608, probably made by Lipperhey,  and
offered it  to  Archduke Albert  (Van Helden, 1977).  Spinola was in The Hague
in that
period of time, as a representant of the Spanish Governor, for peace
negotiations
with Maurice of Nassau,  "Staatshouder" of the seven provinces . 

The presence of the spyglass in the painting can be used to exclude the
year 1608  in the range of the dates for its completion, since the first
spyglasses  appeared in autumn 1608, while  the landscape with the trees
and other details   in the painting  indicates  (late) summer as the
season.

\section{The telescope(s) with draw tubes }

{\it The Allegory of Sight} is one of the series of paintings of  Jan
Bruegel  known as {\it Allegory
of the senses} or {\it The five
senses}, made in
collaboration with  Peter Paul Rubens,   which is conserved at 
the Museum of El Prado, in Madrid.

\begin{figure}
\includegraphics[width=0.9\textwidth]{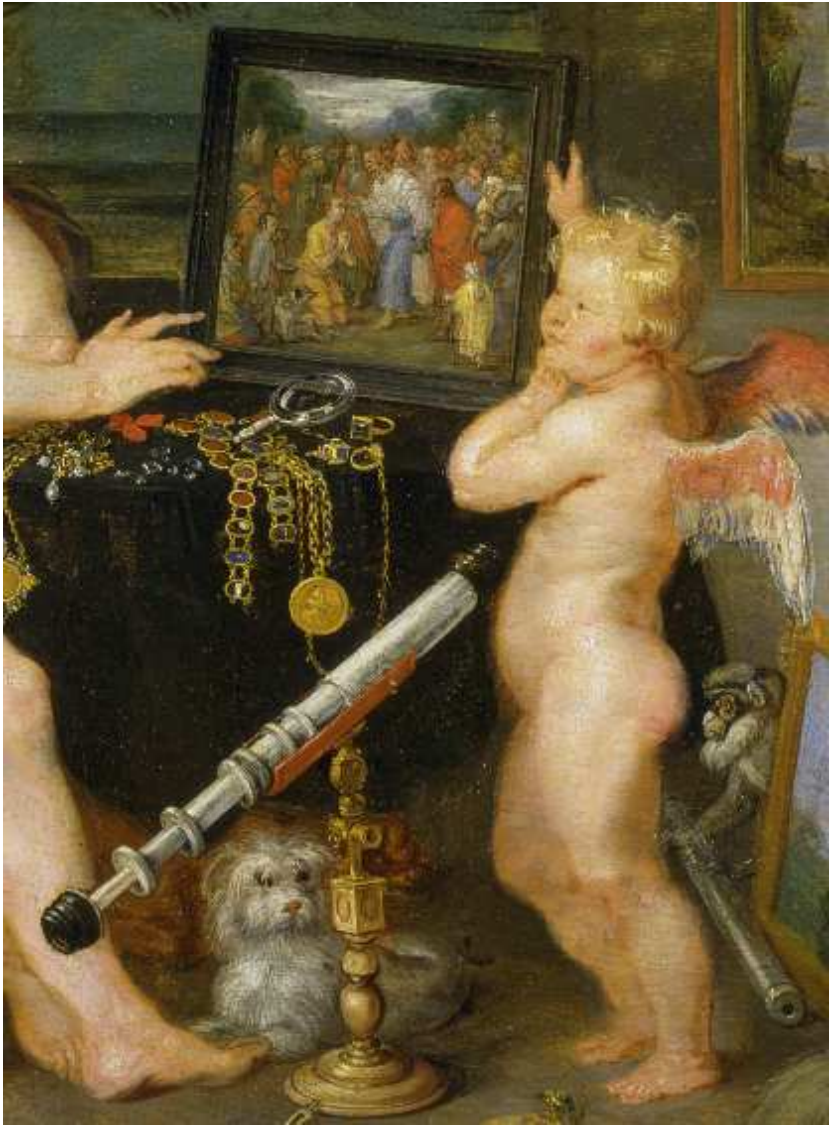}
\caption{A detail of the painting "The Allegory of Sight" by J. Bruegel and
P.P. Rubens,  1617. Madrid, Museo Nacional del Prado. The actual  size of the
telescope  is estimated to be of about 70 cm. } 
\label{fig:1}       
\end{figure}

The painting (oil on wood) is drawn with great accuracy and  depicts a hall
in the ancient royal Palace of
Brussels, on the hill of Coudemberg, residence of the Archdukes,
where their collection of paintings, precious items, and
scientific instruments, most of them related to astronomy, were kept. The
painting had been  completed by  1617 as testified by the
date that appears on a roll of several papers lying over a book entitled 
"Cosmographie", in the
lower part of the painting,
 besides the author's signature.  One can note  various astronomical
instruments such as   a
large astrolabe, an armillary sphere, a pedestal globe, a
proportional compass, map dividers and sundials, which  testify  the
interest of the Archdukes for
science,
and astronomy in particular. Each instrument has been meticulously
characterized with true Flemish skill so that even the minutest details
are accurately reproduced.

The telescope that appears between Venus and Cupid, (Fig. 2)  consists
of a main
tube and seven draw-tubes, all of which appear to be made of metal
(probably silver). Each of the intermediary draw tubes terminates
in an enlarged collar that appears to be made of the same metal.
The lenses are housed in large rounded terminals. 
The instrument is fixed into a curved metal sleeve support attached to a
brass joint which can be adjusted for angle. The pedestal  consists of a
turned column terminating in a simpler saucer-shaped round base.

A comparison with other objects depicted in the painting,  indicates   a
maximum and minimum width for the draw tubes of about 7.5  and 2.5 
cm, respectively, and an estimated  total length, if the tubes were all drawn, 
close
to  170 cm.

It should be  be noted  that a similar telescope is reproduced in a
larger painting, also conserved at the museum "El Prado",   named "The
Allegory
 of the Sight and the Sense of
Smell" (oil on
canvas).  This  painting,  completed around
1618-1620, was commissioned by the City of
Antwerp to J. Brueghel and several other painters (following the 
"kunstkammer" style, that was fashionable at that time), to celebrate the
visit of  the
Archdukes to the city. The  painting,  which actually   is  a copy of the
original
that
was lost in the fire of the Castle of Coudemberg near Bruxelles in 1731,
 includes several instruments
that are very similar to those reproduced in the previous
painting. 

The main difference between  the two telescopes is in the number of
draw tubes (eight instead of seven) and in the color of the rings  (black     
instead of silvery),  see Fig.2 and Fig. 3. Also the pedestals are
different
in the two cases.  Thus, despite the overall similarity in form, which
clearly
indicates  the same origin (maker),  the two telescopes are,  apparently, 
two separate instruments.   We note that Bedini (1971) described this
telescope as constructed with cardboard tubes covered with white or a
light-colored vellum. This seems hardly compatible with the metallic
aspect of the tubes, as suggested by their  color and  reflectivity.

\begin{figure}
  \includegraphics[width=0.9\textwidth]{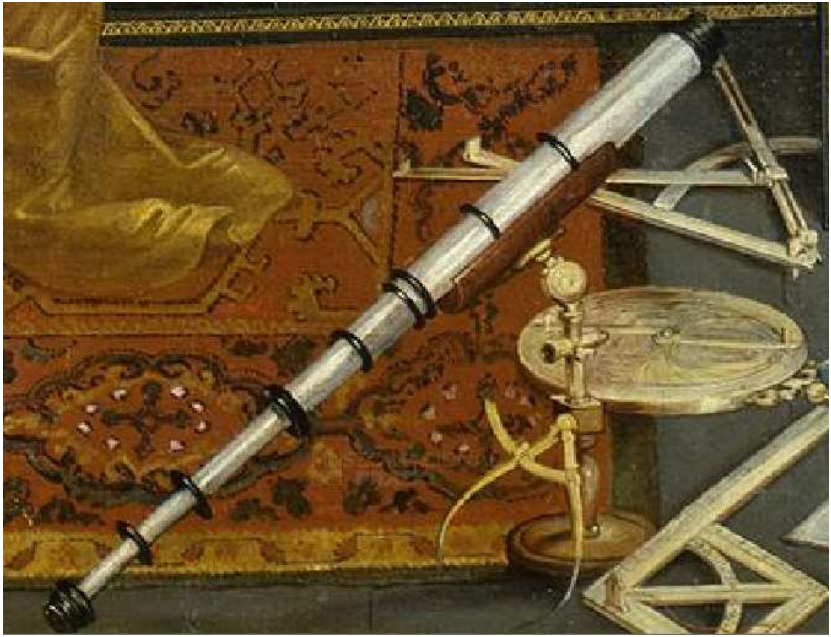}
\caption{A detail of the painting "The Allefory of the Sight and the Sense of
Smell" by J.
Bruegel et al., ca. 1618. Madrid, Museo Nacional del Prado. 
The actual  size of the telescope  in the painting is estimated to be of about
75 cm.}
\label{fig:1}       
\end{figure}

\section{Dutch or keplerian?}

The origin and development of the "astronomical" telescope, the one  
consisting of two convex lenses, is  uncertain and open to questions, like the  
galilean one.  It was theoretically  described by Kepler in his "Dioptrice"
of  
1611 but it is not clear  when the first "astronomical" telescope  was  
manufactured.  "Rosa Ursina", completed by Scheiner in 1630,   is the first
book containing  
a reference to an astronomical telescope.
In 1646, Fontana, in his "Novae Coelestium Terrestriumque Observationes"
claimed  to have manufactured an astronomical 
telescope in 1608, offering also a declaration of father Zupo  
stating  that Fontana had shown a telescope made by two convex lens to  
him and to father Staserio already in   1614.

A definite upper
limit for the manufacturing of the instrument depicted in Fig. 2.  and Fig.
3   is set by the  paintings  completion in  1617  and  1618
respectively. 
What is quite surprising is that their technology appears quite
sophisticated for the epoch and there is no record of similar instruments
until about three decades later (cf. the "Catalogue of Early
Telescopes" by Van Helden, 1999). After examining illustrations of early
telescopes, we conclude that the closest resemblance in shape and  length
is   with the illustrations reported in C. Scheiner's works
({\it Disquisitiones Mathematicae} 1614, and  {\it Rosa Ursina}  of 1631).

We  argue  that these instrument may represent  early
examples of a keplerian telescope, which makes their presence even more
striking.  Three circumstantial
considerations seem to support this view:
\begin{itemize}
\item{ The presence of a quite large terminal (eyepiece) seems to
indicate a
compound eyepiece, quite incompatible with a Galilean
(Dutch) mounting.  In this case, the negative lens needs the eye
to be  brought as close as possible so that the eye's pupil 
becomes the aperture stop and the exit pupil. For the same reason,
if the large terminal were just a lens-screen, this configuration 
would be barely compatible with a Dutch mounting.}

\item{The overall length of the telescope is estimated to be about 180
cm. Even with low powers this would imply a very small field of
view.}

\item{ The first records of Keplerian  telescopes are related to the
Habsburgs.
Cristoph Scheiner in  his "Rosa Ursina" claimed  that he made a keplerian
instrument in 1617 and showed it to the Archduke Maximilian III,
brother of  Albert VII. 
Scheiner,  in a letter on January 4, 1615  wrote 
of a "newly invented instrument"  which actually could be the  
astronomical one, and the date is consistent with the former period. 
According to a different source, Maximilian III, 
about 1615,  received a telescope with two
convex lenses and Scheiner added a third one,  thus manufacturing a
terrestrial keplerian telescope (we note, incidentally, that Scheiner
actually  used a Dutch telescope for his observations of 
sunspots in 1610). Likely,  Albert VII obtained it from Maximilian or heard
of
it and obtained a similar one for his collection.}

\end{itemize}

\section{The monkey's tube}

In the  painting of Fig. 2, on the floor, just behind Cupid, and held by
a monkey,  one can note a tube that "prima facie" looks like an early
spy-glass. Belloni (1964)  described it as an "optical tube", while
Bedini (1971) interpreted it as a microscope,  of the  Janssen
type. 

We recall that  Willem Boreel, born in Middelburg and dutch ambassador in
France, 
 when mentioning that a microscope made by
Janssen
was presented to the Archduke Albert VII, reported a description  made
by Cornelis Drebbel according to whom {\it the microscope had a tube
made of gilded brass, resting on three dolphins made of brass
which were supported by the disc of the base made of ebony} (Van
Helden, 1977). 

A close inspection of a high quality reproduction of the painting
has instead clearly shown that  the tube appears to be made of
 metal (tin or  silver) without any evidence
of the above-mentioned supporting structures.
 The object is
clearly a spyglass and its aspect and size clearly indicates that, most
likely, it corresponds to  the spyglass depicted  in Fig. 1.  

This confirms that  the "tube"   represents 
an early Dutch instrument, as suggested  by Selvelli (1997), on the grounds 
of the  painting of Fig. 2  only.

\begin{acknowledgements}
We gratefully thank Inge Keil and Franz Daxecker for helpful information
and valuable discussion.
\end{acknowledgements}




\end{document}